\documentclass{iopart}
\usepackage{cite}
\usepackage{graphicx}
\usepackage{amsbsy}
\usepackage{amssymb}

\begin{document}

\newcommand{\be}{\begin{equation}}
\newcommand{\ee}{\end{equation}}

\newcommand{\rd}{\mathrm{d}}
\newcommand{\ri}{\mathrm{i}}

\renewcommand{\th}{\tilde{h}}
\renewcommand{\tr}{\tilde{r}}
\newcommand{\tx}{\tilde{x}}
\newcommand{\txi}{\tilde{\xi}}
\newcommand{\tR}{\tilde{R}}
\newcommand{\tRp}{\tilde{R}'}
\newcommand{\tQ}{\tilde{Q}}
\newcommand{\teta}{\tilde{\eta}}
\newcommand{\bep}{\bar\varepsilon}
\newcommand{\tl}{\tilde{\lambda}}

\title[Surface waves and constant vorticity]{How linear surface waves are affected by a current with constant vorticity}

\author{Simen {\AA}. Ellingsen$^1$, Iver Brevik$^1$}

\address{$^1$Department of Energy and Process Engineering, Norwegian University of Science and Technology, N-7491 Trondheim, Norway}
\ead{simen.a.ellingsen@ntnu.no, iver.h.brevik@ntnu.no}

\begin{abstract}
  The interaction of surface waves with Couette-type current with uniform vorticity is a well suited problem for students approaching the theory of surface waves. The problem, although mathematically simple, contains rich physics, and is moreover important in several situations from oceanography and marine technology to microfluidics. We here lay out a simple two-dimensional theory of waves propagating upon a basic flow of uniform vorticity of constant depth. The dispersion relation is found, showing how the shearing current introduces different phase velocities for upstream and downstream propagating waves. The role of surface tension is discussed and applied to the case of a wave pattern created by a moving source, stationary as seen by the source. We conclude by discussing how the average potential and kinetic energies are no longer equal in the presence of shear.
\end{abstract}


\section{Introduction}

Surface waves is one of the main branches of fluid mechanics, treated in many of the defining textbooks in the field, including Lamb \cite{lamb32}, Landau and Lifshitz \cite{landau87} and Chandrasekhar \cite{chandrasekhar61}. Probably the most comprehensive references of the field, incorporating most of what was known about the subject at the time, are the reviews by Wehausen and Laitone \cite{wehausen60}, and Stoker's book \cite{stoker58}. Although many years have passed since their publications, these works continue to be standard references to students, teachers and researchers working on the theory of waves. A string of more recent textbooks have naturally appeared, a majority of which focus on particular aspects or applications, and, as is natural in this day and age, have tended to focus on numerical methods at least as much as analytical. 

A topic which is hardly treated in the standard pedagogical literature to our knowledge, is the interaction of linear surface waves with a shear flow, something which is even true of the more specialised textbooks focussing on waves specifically, such as Lighthill's and Whitham's classical works\cite{lighthill78,whitham74}. An exception is B\"{u}hler \cite{buhler09} whose monograph deals specifically with the interaction of waves and mean flows, and whose framework of analysis is general and powerful and therefore also somewhat advanced. 

In light of its importance (detailed in the next section), the sparsity of analysis of the simplest interaction between shear flow and waves in the standard textbook literature is perhaps surprising. Even more so because, as will be detailed in the following, the problem in its simplest form is not mathematically difficult and provides a number of insights into the interaction of waves and shear flow. Mathemathically tractable while non-trivial, it may be a suitable problem for the intermediate level student being introduced to linear wave theory.

\subsection{Applications}

Water waves on shear flow appear in important applications in oceanography and engineering in two particularly common settings \cite{peregrine76}. Firstly, when wind generates waves on a water surface, the top layer of liquid is set adrift by the wind. Secondly, the shear generated by flow over the bed of a sea or river can manifest itself all the way to the surface when the water is shallow. 

An important branch of applications of the theory of waves running into a shearing current is motivated by the close analogy with waves approaching a beach (e.g.\ \cite{brevik76,brevik79}). As experienced by anyone who have observed waves on the coast, the nature of the ocean waves changes drastically as they approach the shallow region near the shore, the \emph{surf zone}, where the water is a few wavelengths deep and less. As detailed, e.g., by Carter \cite{carter88}, the waves slow up, grow steeper and eventually break, losing much of their energy. Some waves end their lives as so-called \emph{spilling breakers}, gradually breaking over several wavelengths, while others \emph{surge}, i.e., the lower part of the wave surges up on the beach face. The most dramatic breaking is due to \emph{plungers}, waves which break noisily, expending most of their energy in a single splash. 

The model we consider herein is mathematically much easier than a sloping beach but conceptually similar, and serves as a model system where the waves are forced to break, not from shallower water, but from meeting a shear current. Yet applications can be more direct as well, because shear flows can be created for the purpose of wave breaking. For example a sheet of rising bubbles (made, say, by a perforated air pipe under water) creates an up-welling which produces a shear flow near the surface similar to the one we study here, which has the power to ``stop'' approaching waves, forcing them to break \cite{brevik76}. Indeed, in a classical treatment of exactly this phenomenon which had been known (but not understood) for a long time, Evans found that the flow profile near the surface is close to linear \cite{evans55}, and that this current is responsible for calming passing waves. Taylor \cite{taylor55} was able to make use of this linearity of the shear profile to calculate the effect of the current on the flow exactly. The effect of an upflow on passing waves thus serves as a very nice example of how a seemingly very complex phenomenon (the interaction of rising bubbles, current and surface waves) could be understood with a simple theory, in essence that which we present in the following.

Under certain conditions, a shear current can cause the wave group velocity to vanish, creating a ``horizon'' beyond which no waves can propagate, in analogy to event horizons near black holes\cite{nardin09,rousseaux13}. This analogy between surface waves interacting with a current and exotic physics of astronomy is rather attractive, and enables experimentalists to investigate physics which we normally associate with faraway corners of the universe\cite{rousseaux10}. In the other end of the scale, many microfluidic applications involve shear flows in thin liquid layers over surfaces, cf.\ e.g.\ \cite{stone04}. In this case it may not be gravity, but rather surface tension which is the dominating force driving the wave motion. As we shall see, surface tension-driven waves behave quite differently from the gravity waves found in the ocean.

\subsection{Research literature}

While pedagogical works scarecely touch on wave-current interactions, much research has been done on this system, some of which still ongoing. Rousseaux's group have investigated this interaction in recent times \cite{maissa13,maissa13b} with particuar attention to the blocking of waves by a currrent. A similar system was also considered by one of us some time ago, with different collaborators\cite{brevik79,brevik93,brantenberg93}. In these articles waves propagating on a shear flow of uniform vorticity were considered, albeit in a different framework. In the present manuscript we take an arguably simpler approach and show that the problem can be tackled with standard methods. The two-dimensional moving source problem was treated very recently \cite{benzaquen12}, and in the setting of a submerged cylinder in \cite{mccue99}. Waves on a rotational flow emerging from between two plates was considered in \cite{mccue02} where the linearised problem was analysed in detail. The linear waves considered herein are a special case of the more general problem considered by Teles da Silva and Peregrine \cite{telesdasilva88}, and the results discussed herein can mostly be extracted from a study of that reference. Earlier analyses on the same problem from slightly different points of view include in particular \cite{biesel49,thompson50,fenton73,jonsson78,fabrikant98,vanden-broeck00,wahlen07}. Finally we are aware that the closely related system of surface waves generated by a line source, in the presence of constant vorticity, has very recently been considered by Tyvand and Lepper\o d \cite{tyvand13}.

\section{Formulation of the problem}

\begin{figure}[htb]
  \begin{center}\includegraphics[width=3in]{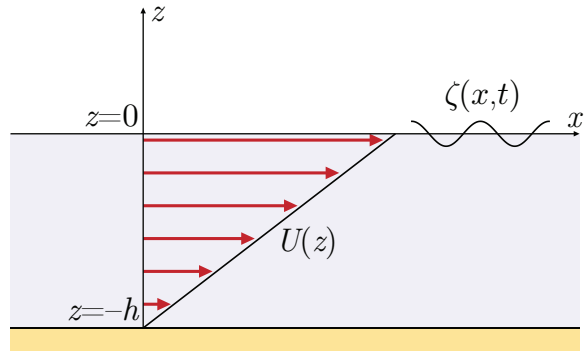}\end{center}
  \caption{The set-up considered: small-amplitude waves propagate on a mean flow with uniform vorticity. When at rest, the surface is at $z=0$ while the bottom is at $z=-h$.}
  \label{fig:geom}
\end{figure}

We consider the geometry shown in figure \ref{fig:geom}. Surface waves described by a presumedly small surface elevation $\eta(x,t)$ are superposed on a mean flow
\be
  U(z) = U_0 + Sz
\ee
where
\be
  S = U_0/h
\ee
is the vorticity, which is constant in space and time. Introducing a separate symbol for the vorticity allows us to keep track of its influence on the solutions as these emerge. When at rest, the surface is at $z=0$ while the bottom is at $z=-h$, so that the depth is constant everywhere and equal to $h$. The density of the fluid is $\rho$, and the surface tension coefficient of the fluid surface towards the atmosphere is $\sigma$.

\subsection{Permissibility of potential theory}

We wish to proceed by adding a small perturbation to the mean flow, and solve the equations of motion to first order in the perturbation when assuming that the additional motion is small on some relevant scale. This is the standard precedure in the study of linear surface waves. The task is made simpler if one is allowed to make use of a scalar velocity potential rather than the vectorial velocity field. One recalls from basic fluid mechanics, however, that the velocity potential normally requires the flow to be irrotational (e.g.\ \cite{landau87} \S9). We argue now, however, that the exception to this rule is when the entire flow is strictly two-dimensional and vorticity is spatially constant.

Consider the vorticity equation (e.g.\ \cite{white06}) which results from taking the curl on both sides of the Navier--Stokes equation:
\be
  \frac{\partial\boldsymbol{\zeta}}{\partial t} = (\boldsymbol{\zeta}\cdot\nabla)\mathbf{v}- (\mathbf{v}\cdot\nabla)\boldsymbol{\zeta}+ \nu\nabla^2 \boldsymbol{\zeta}
\ee
where $\boldsymbol{\zeta}=\nabla\times\mathbf{v}$ is the vorticity and $\mathbf{v}$ the velocity field. Assume that at time $t=0$ the velocity field is strictly two-dimensional, $\mathbf{v}=\mathbf{v}(x,z)$, and that the vorticity is constant, $\boldsymbol{\zeta}=S\hat{\mathbf{y}}$ (we use hats to denote unit vectors). Then all spatial derivatives of $\boldsymbol{\zeta}$ give zero. Moreover, $\mathbf{v}$ depends only on $x$ and $z$ whereas the only nonzero component of  $\boldsymbol{\zeta}$ is in the $y$ direction, so $(\boldsymbol{\zeta}\cdot\nabla)\mathbf{v}=S(\partial/\partial y)\mathbf{v}(x,z)=0$. We are left with $\partial\boldsymbol{\zeta}/\partial t = 0$. In other words, so long as the velocity perturbation we introduce is irrotational and two-dimensional, vorticity is going to \emph{remain} constant for all time. We are thus free to add an a surface wave perturbation in the form of a gradient (i.e., it is irrotational since the curl of a gradient is always zero)
\be
  \mathbf{v}(x,z) = U(z)\hat{\mathbf{x}} + \nabla\phi(x,z,t).
\ee
This observation is most helpful for our purposes. 

\subsection{General solution and boundary conditions}

The local mass conservation relation for incompressible flow, $\nabla\cdot\mathbf{v}=0$, means that the equation of motion for $\phi$ is
\be\label{poisson}
  \nabla^2\phi = \frac{\partial^2\phi}{\partial x^2}+ \frac{\partial^2\phi}{\partial z^2}=0.
\ee
We will find the solution to the equation by means of separation of variables, subject to three boundary conditions. In the following, let $v_x$ and $v_z$ be the $x$ and $z$ components of the velocity field.

The first boundary condition is simply that no fluid can flow through the bottom at $y=-h$, that is,
\be\label{bc1}
  v_z(x,-h)=\frac{\partial\phi}{\partial z}\Bigr|_{z=-h} = 0.
\ee

Let the surface elevation compared to the rest level $z=0$ be $\zeta(x,t)$. 
The second condition, called the \emph{kinematic} boundary condition at the free surface, states that the surface of the liquid is transported with the velocity of the liquid itself. When ignoring higher order terms in small velocities, this means
\[
  \frac{\mathrm{D}\zeta}{\mathrm{D}t} = v_z(x,0)
\] 
with $\mathrm{D}/\mathrm{D}t = \partial/\partial t + (\mathbf{v}\cdot \nabla)$ being the material (or stubstantial) derivative taught in introductory fluid mechanics.
When keeping only linear terms in the small quantities $\zeta$ and $\phi$ we get
\be\label{kin}
  \frac{\partial \zeta}{\partial t}+U_0\frac{\partial \zeta}{\partial x} = \frac{\partial \phi}{\partial z}\Bigr|_{z=0}.
\ee

The third boundary condition is the dynamic one, and states that the pressure just above the interface (i.e., at $z=\zeta$) is constant. To the pressure just below the interface must then be added the pressure jump due to surface tension. Because of the presence of the shear flow this requires a little closer consideration. The $x$ and $z$ components of the Navier--Stokes equations,
\be
  \frac{\partial\mathbf{v}}{\partial t}+(\mathbf{v}\cdot\nabla)\mathbf{v} = -\frac1{\rho}\nabla p - g\hat{\mathbf{z}} + \frac{\mu}{\rho}\nabla^2\mathbf{v}
\ee
($\mu$: dynamic viscosity) read, when again we ignore terms of higher than linear order in small quantities,
\begin{eqnarray}
  \frac{\partial^2\phi}{\partial t\partial x} + U(z)\frac{\partial^2\phi}{\partial x^2} + S\frac{\partial\phi}{\partial z} &=& -\frac1{\rho}\frac{\partial p}{\partial x}; \label{NSx}\\
  \frac{\partial^2\phi}{\partial t\partial z} + U(z)\frac{\partial^2\phi}{\partial x \partial z}  &=& -\frac1{\rho}\frac{\partial p}{\partial z}-g. \label{NSy}
\end{eqnarray}
For this flow, the viscous terms are identically zero. We now integrate (\ref{NSx}) with respect to $x$ and (\ref{NSy}) with respect to $z$, noting that, using partial integration,
\[
  \int\rmd z U(z)\frac{\partial^2\phi}{\partial x\partial z} = U(z)\frac{\partial\phi}{\partial x} - S\int\rmd z \frac{\partial \phi}{\partial x},
\]
and that from $\nabla^2\phi=0$ it follows that
\[
  \int\rmd x\frac{\partial\phi}{\partial z}=-\int\rmd z\frac{\partial\phi}{\partial x} + \mathrm{const}.
\]
Comparing the two, setting $z=\zeta$ and linearising with respect to small quantities gives the dynamical boundary condition, valid at $z=\zeta$,
\be\label{dyn2}
  -\frac{p}{\rho}=\frac{\partial\phi}{\partial t}+U_0\frac{\partial\phi}{\partial x}+S\int\rmd x\frac{\partial\phi}{\partial z}+g\zeta-\frac1{\rho}\Delta p_\mathrm{surf.~tens.}=\mbox{const}.
\ee

When the surface curves, there is a contribution to the pressure just below the surface due to surface tension, which we have included in Eq.~(\ref{dyn2}). We set the atmospheric pressure just \emph{above} the surface to zero (gauge), so that the constant in (\ref{dyn2}), as well as that stemming from the indefinite integral, are zero. 
The pressure jump across the surface due to surface tension has the general form (in 2 dimensions), c.f., e.g., Ref.~\cite{landau87}
\be
  \Delta p_\mathrm{surf.~tens.} = \frac{\sigma}{R}
\ee
where $\sigma$ is the surface tension coefficient, equal to $0.073$N/m for a clean water/air interface, and $R$ is the local radius of curvature of the surface, given in general as \cite{peters01}
\be
  \frac1{R} = \frac{\nabla^2(\zeta-z)}{|\nabla(\zeta-z)|} = \frac{\zeta''}{\sqrt{1+(\zeta')^2}}= \zeta'' + ...
\ee
where $\zeta'=\partial \zeta/\partial x$. Combining everything, the dynamical boundary condition reads
\be\label{dyn}
  \frac{\partial\phi}{\partial t}+U_0\frac{\partial\phi}{\partial x}+S\int\rmd x\frac{\partial\phi}{\partial z}+g\zeta-\frac{\sigma}{\rho}\frac{\partial^2\zeta}{\partial x^2}=0.
\ee
Further details may be found, e.g., in Wehausen and Laitone\cite{wehausen60}.

Assuming separation of variables, we write
\be
  \phi(x,z,t) = X(x,t)Z(z,t).
\ee
Proceeding in the standard procedure taught in calculus courses, the solution to governing equation (\ref{poisson}), when subjected to boundary condition (\ref{bc1}), is
\be
  \phi(x,z,t) = A(k,t) \rme^{ikx}\cosh k(z+h)
\ee
for some arbitrary wave number $k\geq 0$. Naturally, the physical quantity is the real part. A wave number $k$ corresponds to a wavelength
\be
  \lambda = \frac{2\pi}{k}.
\ee

Since the wavelength of the surface deformation must equal that of the velocity field we may likewise write
\be
  \zeta(x,t) = B(k,t)\rme^{ikx}.
\ee

\section{Dispersion relation for wave motion}\label{sec:disp}

We are now ready to insert the general solutions into the boundary conditions (\ref{kin}) and (\ref{dyn}) to solve the propagating motion of the surface wave solutions. To simplify our analysis a little, let us leave out the effect of surface tension for now (we come back to it later). As a rule of thumb, waves on water in the centimeter regime and larger are not much affected by surface tension, so for water waves of this size and above, this simplification is a good approximation. Surface tension is important, however, for millimeter waves and smaller, where it is a driving force for the so-called capillary waves. We return to this in section \ref{sec:capillary}.

Leaving the surface tension term out, equations (\ref{kin}) and (\ref{dyn}) give, respectively,
\begin{eqnarray}
  \dot B + ikU_0 B -kA \sinh kh&=&0  \label{eA}\\
  (\dot A + ikU_0 A)\cosh kh - iS A\sinh kh + gB &=& 0.\label{eB}
\end{eqnarray}
A dot here denotes differentiation w.r.t.\ time. Eliminating $A$ using Eq.~(\ref{eA}) and inserting into Eq.~(\ref{eB}) gives the transient equation for $B$ which may be written
\be
  \ddot B + 2i\omega_1\dot B + \omega_2^2B=0
\ee
where we have defined the frequencies
\begin{eqnarray}
  \omega_1&=& kU_0 - \frac1{2}S\tanh kh \\
  \omega_2^2&=& k(g+SU_0)\tanh kh - k^2 U_0^2.
\end{eqnarray}
The solutions are
\be\label{B1}
  B(k,t) = \beta_u(k)\rme^{-i(\omega_1-\sqrt{\omega_1^2+\omega_2^2})t} + \beta_d(k)\rme^{-i(\omega_1+\sqrt{\omega_1^2+\omega_2^2})t}.
\ee
The constants $\beta_u$ and $\beta_d$ are still undetermined amplitudes, and must be determined from initial conditions. Subscripts $u$ and $d$ denote ``upstream'' and ``downstream'' for reasons which will become clear shortly.
We shall not be concerned with details of $\beta_u$ and $\beta_d$ in the following, as we do not wish to specify a particular source for the waves, and simply take them to be known quantities.

A surface wave propagating in the $x$ direction at phase velocity $c$ has the form $\rme^{ik(x-ct)}$, so with equation (\ref{B1}) for $B$ the phase velocities of propagating waves have been found. We choose now to take the surface of the water as the frame of reference when considering the wave velocities. This is natural, for example, if the application we have in mind is a boat (or other wave source) moving on the surface. We find the propagation speed $c$ relative to the motion of the liquid surface by excluding from its definition the constant surface motion $U_0$ and write $\zeta$ with Eq.~(\ref{B1}) as
\be
  \zeta(x,t) = \beta_u(k) \rme^{ik(x-U_0 t + c_u t)}+\beta_d(k) \rme^{ik(x-U_0 t - c_d t)}.
\ee
We have replaced $\omega_1$ and $\omega_2$ by an \emph{upstream} and a \emph{downstream} phase velocity relative to the liquid surface, which we can write
\begin{eqnarray}
  c_u(k) &=& \sqrt{\frac{g}{k}\tanh kh + \Bigl(\frac{S}{2k}\tanh kh\Bigr)^2}+\frac{S}{2k}\tanh kh,\\
  c_d(k) &=& \sqrt{\frac{g}{k}\tanh kh + \Bigl(\frac{S}{2k}\tanh kh\Bigr)^2}-\frac{S}{2k}\tanh kh.
\end{eqnarray}

For analysis it is instructive to write the velocities in the following form:
\begin{eqnarray}
  c_u(k) &=& \sqrt{gh}\Bigl[\sqrt{\frac{\tanh kh}{kh} + \Bigl(\mathrm{Fr}_h\frac{\tanh kh}{kh}\Bigr)^2}+\mathrm{Fr}_h\frac{\tanh kh}{kh}\Bigr],\\
  c_d(k) &=& \sqrt{gh}\Bigl[\sqrt{\frac{\tanh kh}{kh} + \Bigl(\mathrm{Fr}_h\frac{\tanh kh}{kh}\Bigr)^2}-\mathrm{Fr}_h\frac{\tanh kh}{kh}\Bigr],\\
  \mathrm{Fr}_h &=& \frac{S\sqrt{h}}{2\sqrt{g}} = \frac{U_0}{2\sqrt{gh}}.\label{Frh}
\end{eqnarray}
The dimensionless number $\mathrm{Fr}_h$ we may call the shear Froude number. It is a measure of the amount of shear (or vorticity) which is present. This dispersion relation is the same as may be found in Refs.~\cite{fenton73,telesdasilva88} and perhaps most closely to the present notation, \cite{jonsson78, maissa13,maissa13b}.

\begin{figure}[htb]
  \begin{center}
    \includegraphics[width=.7\textwidth]{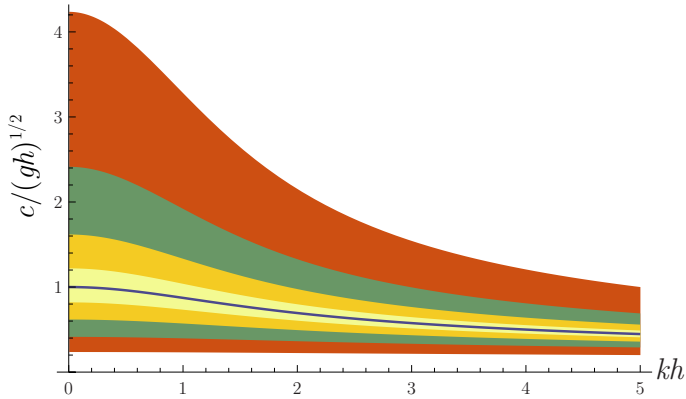} 
    \caption{Splitting of upstream and downstream phase velocities due to the presence of shear $S$. Phase velocity as function of $kh$ are plotted for different values of $\mathrm{Fr}_h$, divided by $\sqrt{gh}$. The central curve is $\mathrm{Fr}_h=0$ (zero shear) when velocities are equal. The shaded regions outside are bounded above by the upstream velocity and below by downstream velocity at increasing $\mathrm{Fr}_h$ (innermost to outermost region): $\mathrm{Fr}_h=0.2,0.5,1$ and $2$.}
    \label{fig:c}
  \end{center}
\end{figure}

In the limit $\mathrm{Fr}_h \to 0$ (that is, zero shear), the well known dispersion relation for pure gravity waves is regained (c.f., e.g., Ref.~\cite{wehausen60}),
\be
  c_u=c_d = \sqrt{\frac{g}{k}\tanh kh}.
\ee
We then see that the role played by the shear is a \emph{splitting of phase velocities}. Upstream propagating waves move more quickly because of the shear, while downstream propagating ones are slower. This is illustrated in figure \ref{fig:c}, where the upstream an downtream phase velocities are plotted for different $kh$ at four different Froude numbers $\mathrm{Fr}_h$.

\subsection{Group velocity}

Knowing the dispersion relations $c(k)$ for the phase velocity, it is straightforward to calculate the corresponding group velocities,
\be
  c_{g,u}(k) = \frac{\rmd  }{\rmd k}[kc_u(k)]; ~~~ c_{g,d}(k) = \frac{\rmd  }{\rmd k}[kc_d(k)].
\ee
The resulting expressions are somewhat bulky, and we do not quote them explicitly. 

\begin{figure}[htb]
  \begin{center}
    \includegraphics[width=.7\textwidth]{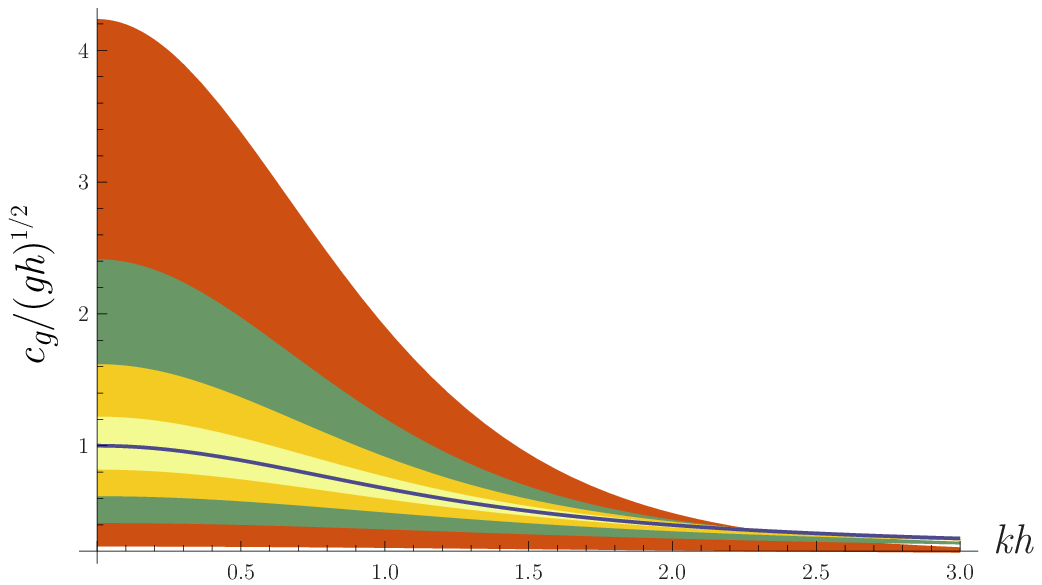} \\
    \includegraphics[width=.7\textwidth]{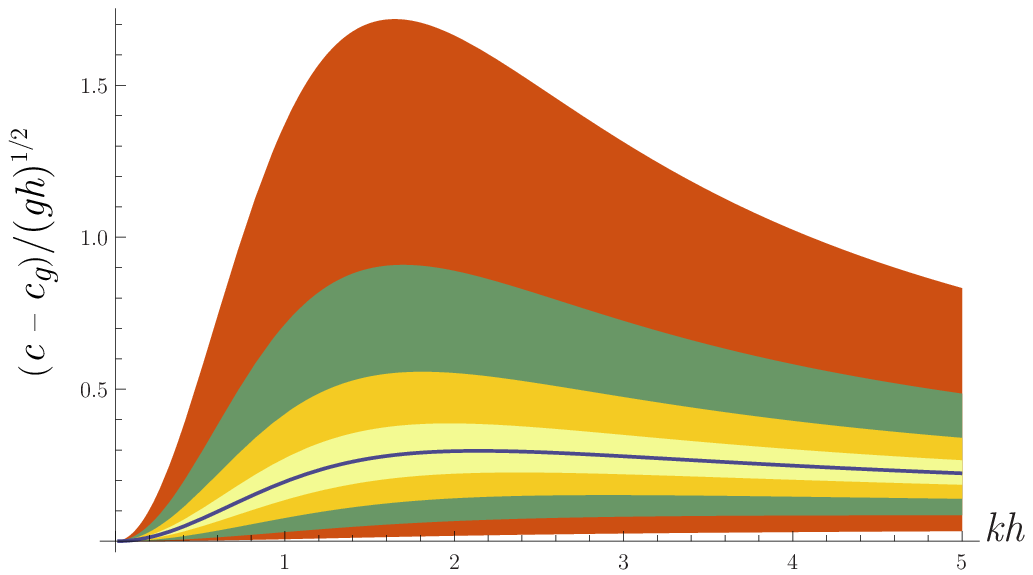} 
    \caption{Top: Same as figure \ref{fig:c}, but for the group velocity $c_g=\rmd [kc(k)]/\rmd k$. Bottom: difference $c-c_g$ for upstream and downstream wave modes.}
    \label{fig:cg}
  \end{center}
\end{figure}

Just as for the phase velocity, the presence of shear $S>0$ makes for a splitting into different upstream and downstream group velocities. Group velocities are plotted as function of $kh$ in figure \ref{fig:cg} for the same Froude numbers as in figure \ref{fig:c}. When the shear Froude number is of order $1$ we see that $c_{g,u}\gg c_{g,d}$ for all but the very shortest wavelengths. In the lower panel of figure \ref{fig:cg}, the difference between phase velocity and group velocity is shown for our gravity waves. In this case, the difference is always positive: waves driven by gravity alone have group velocity which is slower than the phase velocity. This changes when also surface tension is considered.

\subsection{Wave stopping}

How does our waves look when observed not by someone following the surface flow, but rather by someone standing on shore? A group of waves moving downstream now seems to move at phase velocity $U_0 + c_{g,d}$ while an upstream moving group moves at velocity $U_0 - c_{g,u}$. If now $U_0>c_{g,u}$, no group of waves can move upstream relative to the bottom of the flow. This is called \emph{wave stopping}, and has direct applications (e.g.\ \cite{brevik76,evans55,taylor55} as well as interesting analogies to black holes \cite{nardin09,rousseaux13}. Of course, wave stopping does not require the presence of shear; in fact we have seen that the shear \emph{accelerates} the upstream propagating waves, so a uniform current towards positive $x$ would in fact have a higher stopping power. In pratice, however, setting up a wave breaking current entails that shear is present, as illustrated by the bubble curtain example \cite{evans55,taylor55} which we discussed in the introduction.

\section{The role of surface tension}\label{sec:capillary}

Let us now include the surface tension term in the dynamic boundary condition (\ref{dyn}). Equation (\ref{eB}) then reads
\[
  (\dot A + ikU_0 A)\cosh kh - iS A\sinh kh + (g+k^2\sigma/\rho)B = 0
\]
so in all our previous results, including surface tension amounts to replaing $g$ by $g+k^2\sigma/\rho$. 
The phase velocities are then modified to
\begin{eqnarray}
  c_u(k) &=& \sqrt{\Bigl(\frac{g}{k}+\frac{\sigma k}{\rho}\Bigr)\tanh kh + \Bigl(\frac{S}{2k}\tanh kh\Bigr)^2}+\frac{S}{2k}\tanh kh,\label{cu}\\
  c_d(k) &=& \sqrt{\Bigl(\frac{g}{k}+\frac{\sigma k}{\rho}\Bigr)\tanh kh + \Bigl(\frac{S}{2k}\tanh kh\Bigr)^2}-\frac{S}{2k}\tanh kh.\label{cd}
\end{eqnarray}
Again we recognise the $S\to 0$ limit as the classic result \cite{wehausen60}
\be
  c_u(k)=c_d(k) = \sqrt{\Bigl(\frac{g}{k}+\frac{\sigma k}{\rho}\Bigr)\tanh kh}.
\ee
Thus there is no further coupling between surface tension and the shear flow. However, the inclusion of $\sigma$ has the effect it always has, namely to make sure that there exists a \emph{minimum phase velocity} where $\rmd c(k)/\rmd k=0$, and that for a velocity $V$ greater than the minimum velocity the equation $c(k)=V$ has two solutions of which one is a wave primarily driven by gravitation and the other by capillary (i.e., surface tension) forces. This becomes important when we next consider the case of a travelling wave source. 

\section{Waves from a travelling source}

We will now concentrate on wave generated by a pressure line source travelling along the $x$ axis at velocity $V$ relative to the water surface ($V + U_0$ relative to sea bed and coordinate system). We choose by convention that $S$ (and thus also $U_0$) is positive, and that $V>0$ represents downstream motion and $V<0$ upstream motion. This system was considered very recently by Benzaquen and Rapha\"{e}l \cite{benzaquen12}. This is a two-dimensional model version of the waves produced behind a boat, a problem first considered by Kelvin a long time ago \cite{kelvin1887} and discussed by Lamb \cite{lamb32}. 

Following Ref.~\cite{raphael96}, we assume that the surface perturbations appear stationary when beheld by the travelling source, that is, the surface waves depend not on $x$ and $t$ independently, but only on the combination $\xi = x-(U_0+V)t$. Comparing with our previous solutions and dispersion relations this means that for wave mode of wave number $k$,
\be
  V = \left\{\begin{array}{cl}-c_u(k),   &\mbox{when } V<0 \mbox{ (upstream)},\\c_d(k), &\mbox{when }V>0\mbox{ (downstream)}.\end{array}\right.
\ee
In order to limit the number of parameters, we will consider the deep and shallow water limits only. 

\subsection{Deep water}

In the deep water limit, $kh\to \infty$, the phase velocities read,
\begin{eqnarray}
  c_{{}^u_d}(k) &\buildrel{kh\to\infty}\over{=}& \sqrt{c_0^2(k) + \Bigl(\frac{S}{2k}\Bigr)^2}\pm\frac{S}{2k} \\
  c_0^2(k)&=&\frac{g}{k}+\frac{k\sigma}{\rho}\label{c0}
\end{eqnarray}
[compact notation: upper (lower) letter $u~(v)$ goes with upper (lower) sign as in equations (\ref{cu}) and (\ref{cd})].

\begin{figure}[htb]
  \begin{center}
    \includegraphics[width=.9\textwidth]{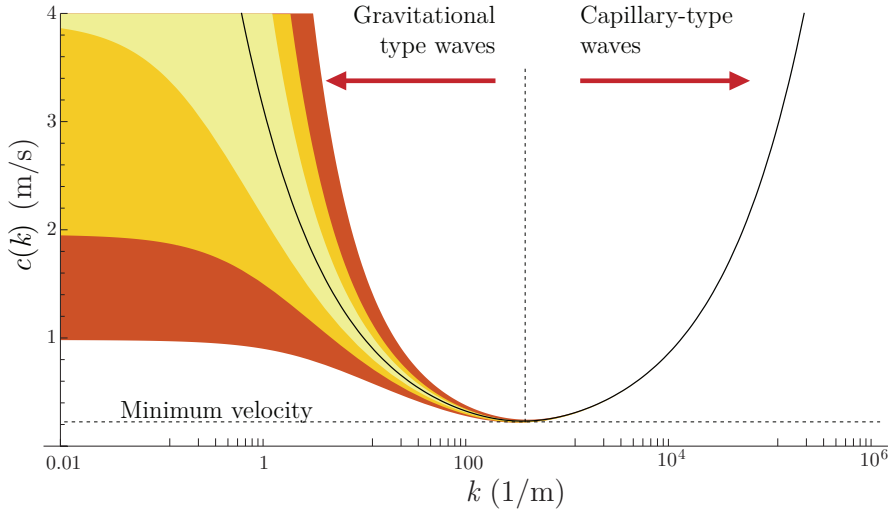} 
    \caption{Phase velocities as function of wave number $k$ in deep water. The shaded areas are bounded below by $c_d(k)$ and above by $c_u(k)$ for three cases from lighter to darker: $S=2.5$/s, $5$/s and $10$/s.}
    \label{fig:deep}
  \end{center}
\end{figure}

In figure \ref{fig:deep} we plot the phase velocities for upstream and downstream propagation for the case of water: $\rho=997$\,kg/m$^3$, $\sigma = 0.073$\,N/m. The first thing we notice is that there exists a minimum propagation velocity. At zero shear this is about $23$\,cm/s for water. If the source moves more slowly than this, there exist no linear waves which solve $V=c$. In the presence of shear, the minimum velocity as well as the wave number at which it is found, are sligthly lowered for downstream waves and slightly increased for upstream waves, but this is only noticeable for extremely strong shear. 

For any source velocity $V$ greater than the miminum value, there are two wave numbers which solve $c(k)=|V|$ (either upstream or downstream phase velocity as appropriate). The solution corresponding to the smaller $k$ (long wavelength) we call \emph{gravity-type}, since it is the only solution obtained in the absence of surface tension. The larger $k$ solution (short wavelength) we call \emph{capillary type}, since it is a wave driven primarily by surface tension.

The next thing to notice is that when the source travels much faster than the minimum velocity of $23$\,cm/s, the wavelength of the capillary solution grows very small. Already at $2$\,m/s the capillary water waves have wavelength of about a tenth of a millimeter. Hence for sources moving in meters per second, the capillary waves will hardly be visible to the naked eye and do not contribute noticeably to loss of energy.

Now note that since $c_g(k)=\rmd[ kc(k)]/\rmd k = c(k)+kc'(k)$, the difference between group velocity and phase velocity is always 
\be
  c_g(k) -c(k) = k c'(k).
\ee
At the minimum phase velocity, $c'(k)=0$, so upon inspection we conclude that $c_g>c$ for capillary waves, and $c_g<c$ for gravity waves. Thus, since wave energy is propagated at the group velocity, gravity waves fall behind the moving source, while capillary waves travel ahead of the source.

The third observation we make from figure \ref{fig:deep} is that the presence of shear matters very much for the gravity type waves and virtually not at all for the capillary waves. This is perhaps not so surprising: the capillary waves are very small and hardly penetrate into the interior of the fluid body, hence do not care much what other flow might be present there. Not so for the longer-wavelength gravity waves, however. In particular in the limit of very long waves, the downstream phase velocity no longer becomes arbitrarily large, but tends to a constant:
\be
  c_d(k)\buildrel{k\to 0}\over{\longrightarrow} g/S = c_{\mbox{\scriptsize lim}}.
\ee
Thus, a wave source travelling downstream at $V>c_{\mbox{\scriptsize lim}}$ can only generate stationary capillary waves, not gravity-type waves. 

We could speak of $c_{\mbox{\scriptsize lim}}$ being a \emph{critical velocity} for downstream motion. A very similar situation exists for boats travelling over shallow water, where (as we shall explain shortly) the phase speed is also bounded by a maximum velocity $\sqrt{gh}$ in the long wavelength limit. When the boat moves faster than $\sqrt{gh}$, no waves can follow it while propagating parallel to the direction of motion. The only waves generated then propagate at almost normal angle away from the wake, while directly behind the boat, where \emph{transverse} waves following the boat are normally found, the wake is free of waves. A classic discussion of this is due to Havelock \cite{havelock08}.

\subsection{Shallow water}

\begin{figure}[htb]
  \begin{center}
    \includegraphics[width=.8\textwidth]{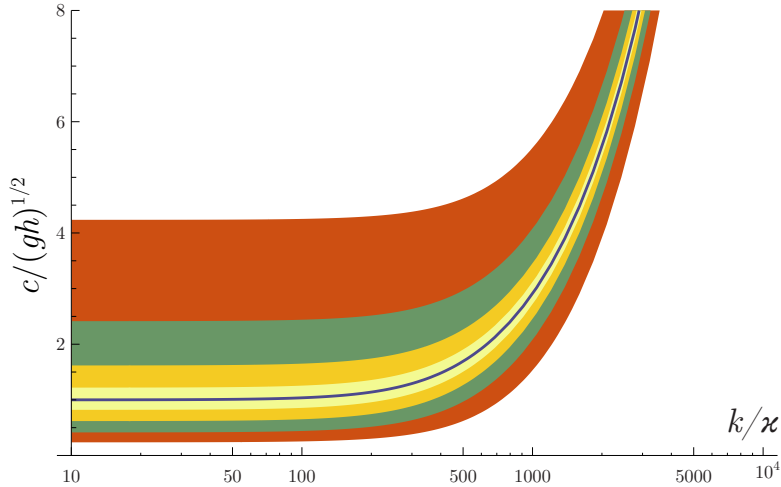} 
    \caption{Phase velocities as function of $k/\varkappa$ in shallow water. The shaded areas are bounded below by $c_d(k)$ and above by $c_u(k)$ for three cases from lighter to darker: $\mathrm{Fr}_h=0.2,0.5,1$ and $2$.}
    \label{fig:shallow}
  \end{center}
\end{figure}

In the shallow water limit, $kh\ll 1$, the phase velocities can be written conveniently,
\be
  c_{{}^u_d}(k) \buildrel{kh\ll 1}\over{=} c_0\Bigl[\sqrt{1 + \Bigl(\frac{\mathrm{Fr}_h}{c_0/\sqrt{gh}}\Bigr)^2}\pm\frac{\mathrm{Fr}_h}{c_0/\sqrt{gh}}\Bigr]
\ee
where $\mathrm{Fr}_h$ was defined in equation (\ref{Frh}) and $c_0$ in equation (\ref{c0}), and 
\be
  \frac{c_0(k)}{\sqrt{gh}} = \sqrt{1+\frac{k^2}{\varkappa^2}}; ~~ \varkappa = \sqrt{\frac{\rho g}{\sigma}}\approx 366\mbox{m}^{-1}
\ee
is the velocity in the absence of shear. $\varkappa$ is sometimes called the capillary wave number and the numerical value given is for water. We plot the shallow water phase velocities in figure \ref{fig:shallow}.

In shallow water the situation is very different from the deep-water limit. The minimum velocity is found in the limit $k\to 0$, and is 
\be
  c_{{}^u_d}(k) \to \sqrt{gh}\Bigl[\sqrt{1+\mathrm{Fr}_h^2} \pm \mathrm{Fr}_h\Bigr].
\ee
For any velocity $V$ smaller than the minimum velocity (upstream or downstream as appropriate), no steady solution exists. Above the minimum velocity, there is now only one solution to $c(k)=V$. This solution is of capillary type. We see this from noticing that whenever a solutions exists, it lies at $k\gtrsim 200\varkappa$. The capillary wavelength $\lambda_c=2\pi/\varkappa$ is only about $1.7$\,cm, so the stationary wave solution has a wavelength two orders of magnitude shorter than this. 

It is important to note, however, the ``shallow water limit'' means, namely that the water depth is much smaller than a wavelength. When we then observe that after taking this limit, the only solution to $V=c$ has extremely short wavelength, one must of course check if the shallow water approximation still holds. A typical situation at finite depth is that the longer wavelengths ``see'' the water as shallow, whereas the short wavelength capillary waves still ``see'' deep waters. The above considerations provide all the tools for considering this more general situation, but we shall not go further into it here.

Another important observation is that when shear Froude number is large, stationary solutions for sources travelling upstream exist only for sources traveling significantly faster than $\sqrt{gh}$. A source travelling more slowly cannot keep up with the waves it creates, and the wave picture can never be stationary as seen by the source.

\section{Critical layers}

When considering waves in a more general shear flow $U(z)$, a \emph{critical layer} appears at a depth $z_c$ if at that depth $U(z_c)=c_u(k)$. A thorough treatment of this phenomenon is found in a classical paper by Booker and Bretherton \cite{booker67}. In linear wave theory of this more general type the vertical velocity component $v_z$ satisfies the equation (for incompressible flow) \cite{booker67,leblond78}
\be
  \frac{\partial^2 v_z}{\partial z^2}-\Bigl[\frac{U''(z)}{U(z)-c_u(k)} + k^2\Bigr]v_z = 0
\ee
which is called the Rayleigh equation.
Four our simple Couette profile, $U''(z)=0$, so the fraction disappears. If the mean velocity profile curves, however, the equation becomes singular at $z=z_c$, and the mathematics near the critical layer become quite involved \cite {miles61}. The dynamics of critical layers plays an important role in the transfer of energy from wind to ocean waves\cite{miles57}, and energy transfer by such mechanisms is directly proportional to $U''(z_c)$. It is important to notice, thus, that a model situation with a constant shear will fail to capture such nonlinear phenomena. Critical layers are discussed in a very readable fashion in chapter 7 of \cite{leblond78}. In particular, Lord Kelvin showed that characteristic ``cat's eye'' vortices form near such a layer \cite{leblond78,kelvin1880}.

\section{Mean mechanical energy}

In the absence of shear, the mechanical energy integrated over one wavelength of a linear wave is equally divided between kinetic and potential energy (e.g.\ \cite{landau87}). Indeed, this is true of unforced harmonic oscillators in general, and is the famous \emph{principle of equipartition of energy} from classical mechanics. We shall see that when $S\neq 0$, this is not true of the surface waves we have considered herein, a point considered in further depth (albeit in a much more involved formalism) in \cite{telesdasilva88}.

For the purposes of energy, it is useful for us to use a real rather than complex formalism. Taking the real part of our previous complex surface elevation, we can write it, as seen relative to the motion of the surface of the water, as
\be
  \zeta(x,t) = \beta(k)\cos k \xi
\ee
where we now assume $\beta$ to be real and we use the shorthand
\be
  \xi = x- U_0 t - ct
\ee
where $c$ is the phase velocity. For a downstream propagating wave, $c = c_d$, and for upstream, $c=-c_u$, with expressions as given in equations (\ref{cu}) and (\ref{cd}) before. In this section we will again ignore surface tension for simplicity.

We repeat the procedure of separation of variables we introduced in chapter \ref{sec:disp} and subject it to the requirement of $v_z=0$ at $z=-h$, and obtain in this case
\be\label{phiE}
  \phi(x,z,t)=-c\beta(k)\frac{\cosh k(z+h)}{\sinh kh}\sin k\xi.
\ee

The potential energy integrated over one wavelength $\lambda = 2\pi/k$ is
\be
  \mathrm{PE} = \int_0^\lambda \rmd \xi\int_{-h}^\zeta\rmd z  \rho g z = \frac1{2}\rho g \int_0^\lambda \rmd \xi(\zeta^2-h^2).
\ee
We are only interested in the part of the energy associated with the waves, so the term $h^2$ is neglected, leaving the \emph{additional} potential energy due to the wave motion:
\be
  \mathrm{PE}_w = \frac1{2}\rho g \int_0^\lambda \rmd \xi\zeta^2 = \frac{\pi}{2}\rho\beta^2\frac{g}{k}.
\ee

Next, the kinetic energy is found by integrating the kinetic energy density
\be
  \frac1{2}\rho |\mathbf{v}|^2 = \frac1{2}\rho \left[U^2(z) + 2U(z)\frac{\partial \phi}{\partial x} + \left(\frac{\partial \phi}{\partial x}\right)^2+\left(\frac{\partial \phi}{\partial z}\right)^2\right].
\ee
Again we neglect the term $\propto U^2$ since it concerns the basic flow only. Next, the term $\propto U$ integrates to zero over a wavelength because $\partial \phi/\partial x \propto \cos kx$. We are left with the wave kinetic energy
\be
  \mathrm{KE}_w = \frac1{2}\rho g \int_{-h}^\zeta\rmd z\int_0^\lambda \rmd \xi\left[\left(\frac{\partial \phi}{\partial x}\right)^2+\left(\frac{\partial \phi}{\partial z}\right)^2\right] = \frac{\pi\rho\beta^2c^2}{2\tanh^2kh}.
\ee
The details of inserting $\phi$ from Eq.~(\ref{phiE}), differentiating, and solving the last integral are left to the student.

We now see that $\mathrm{KE}_w=\mathrm{PE}_w$ only if
\[
  c = \pm \sqrt{\frac{g}{k}}\tanh kh.
\]
This is true when $S=0$, but not if $S>0 $. To wit, in the presence of shear, $\mathrm{KE}_w>\mathrm{PE}_w$ upstream propagating waves, and $\mathrm{KE}_w<\mathrm{PE}_w$ for downstream propagating waves. 

This fact has some important implications, because in practice wave energy spectra are often produced based on measuring surface elevation only, using the equipartition principle to determine mechanical energy \cite{telesdasilva88}. This procedure could be misleading in the presence of shear, unless care is taken.

\section{Conclusions}

The interaction of linear waves with a uniform shear flow is a simple system with rich physics, touching on many of the most central concepts in wave theory. As such it is a system well suited for an intermediate level student approaching the subject, yet little textbook material exists at the appropriate level. In the present paper we give an exposition of the system of surface waves interacting with uniform vorticity in a mathematically simple framework, bringing forth many of the interesting physical properties in a straightforward manner.

Particular interest is given to the dispersion relation, i.e., the dependence of a wave's phase velocity on its wavelength. The role of vorticity compared to waves on a uniform flow is to introduce different phase velocities for waves moving upstream and downstream: upstream waves move faster and downstream ones slower than in the absence of shear. The limits of deep and shallow waters are considered separately, with differing behaviour in each case. 

When waves are assumed stationary as seen by a moving wave source, the deep and shallow cases behave very differently. In deep waters the presence of surface tension means there exists a minimum velocity below which no wave can propagate. For sources moving faster than the minimum velocity, two types of waves can be generated: long wavelength ``gravity type'' waves falling behind the source, and short wavelength ``capillary type'' waves moving ahead of the source. When the source moves much faster than this minimum velocity, $23$\,cm/s for water, the capillary waves are negligible. In shallow water the situation is quite different: the minimum velocity is found in the limit of infinite wavelength and increases monotonously thence for decreasing wavelength. For a source moving more slowly than the minimum, no stationary waves exist; all waves move faster than the source. A source moving faster than the minimum will only generate capillary type waves of very short wavelength. 

As a final point we discuss the division of mean mechanical energy into potential and kinetic energy. In the absence of the shear flow the principle of equipartition holds, and the potential and kinetic energy are the same on average. However, because the phase velocity for upstream vs downstream propagating waves each differ from the no-shear case, we show that in the present case, mean kinetic energy is greater than mean potential energy for upstream propagating waves, and the opposite for downstream waves. 

\section*{Acknowledgements}

We have benefited from discussions with Prof.\ Peder A.\ Tyvand in the preparation of this article. We are thankful to Prof.~Germain Rousseaux for helpful comments, as well as to E.~Rapha\"{e}l, S.~W.\ McCue and Y.~A.\ Stepanyants for valuable input.

\end{document}